\documentclass{article}

\usepackage[final,nonatbib]{neurips_2020_ml4ps}

\usepackage[utf8]{inputenc} % allow utf-8 input
\usepackage[T1]{fontenc}    % use 8-bit T1 fonts
\usepackage{hyperref}       % hyperlinks
\usepackage{url}            % simple URL typesetting
\usepackage{booktabs}       % professional-quality tables
\usepackage{amsfonts}       % blackboard math symbols
\usepackage{nicefrac}       % compact symbols for 1/2, etc.
\usepackage{microtype}      % microtypography
\usepackage{amsmath}
\usepackage{graphicx}
\usepackage{multicol}
\usepackage[dvipsnames,svgnames,x11names]{xcolor}
\usepackage{bm}
\usepackage[export]{adjustbox}
\usepackage[caption = false]{subfig}

\newcommand{\vq}{{\mathbf{q}}}
\newcommand{\vx}{{\mathbf{x}}}
\newcommand{\vPsi}{{\bm{\Psi}}}
\newcommand{\vk}{{\mathbf{k}}}

\title{Fast and Accurate Non-Linear Predictions of Universes with Deep Learning}

\author{
  Renan Alves de Oliveira\\
  PPGCosmo/Flatiron Institute\\
  \texttt{fisica.renan@gmail.com} \\
  \And
  Yin Li\\
  Flatiron Institute\\
  \texttt{yinli@flatironinstitute.org} \\
  \And
  Francisco Villaescusa-Navarro\\
  Princeton University\\
  \texttt{villaescusa.francisco@gmail.com} \\
  \And
  Shirley Ho\\
  Flatiron Institute\\
  \texttt{shirleysurelyho@gmail.com} \\
  \And
  David N. Spergel\\
  Flatiron Institute\\
  \texttt{dspergel@flatironinstitute.org} \\
}

\begin{document}

\maketitle

\begin{abstract}
Cosmologists aim to model the evolution of initially low amplitude Gaussian density fluctuations into the highly non-linear "cosmic web" of galaxies and clusters. They aim to compare simulations of this structure formation process with observations of large-scale structure traced by galaxies and infer the properties of the dark energy and dark matter that make up 95\% of the universe. These ensembles of simulations of billions of galaxies are computationally demanding, so that more efficient approaches to tracing the non-linear growth of structure are needed. We build a V-Net based model that transforms fast linear predictions into fully nonlinear predictions from numerical simulations. Our NN model learns to emulate the simulations down to small scales and is both faster and more accurate than the current state-of-the-art approximate methods.
It also achieves comparable accuracy when tested on universes of significantly different cosmological parameters from the one used in training. This suggests that our model generalizes well beyond our training set. 
\end{abstract}

\section{Introduction}

Cosmology is at critical stage: the 2019 Nobel Prize recognized that a simple model with only five basic parameters fits a host of astronomical observations on large-scales where the fluctuations are in the linear regime. This model implies that atoms make up only 5\% of the universe with the remaining 95\% in the form of dark matter and dark energy. Astronomers have embarked on several large-scale surveys and are launching multiple satellites that aim to collect data to understand these mysteries. Analysis of this new data is limited by our computational abilities as numerical simulations are essential for comparing theoretical predictions to observations on the small non-linear scales that contain most of the cosmological information~\cite{Chang_19, Massara_2020, Cora_19, Allys_2020, Arka_Tom_2020}. Unfortunately, the computational cost of these simulations is high; a single simulation may require from thousands to millions of CPU hours.

In this work, we aim to build a neural network (NN) model that can \textit{learn} to emulate these numerical simulations both accurately and efficiently.
This paper builds on the work of He et al.~\cite{He13825}, which showed that neural networks can indeed learn to run \emph{fast approximate} simulations.
Our main contributions is improvement of the model by
training on \emph{full N-body} simulations and extending to much smaller scales,
where the learning task becomes more difficult due to nonlinearity.

\section{Methods}

\paragraph{N-body Simulation.}

An $N$-body simulation evolves a large number of massive dark matter particles interacting with each other only through Newtonian gravity.
Since our Universe began with matter distributed almost uniformly, the simulation starts with particles only slightly perturbed from a uniform grid and nearly at rest.
During the simulation, a particle moves from its initial location $\vq$ to its final position $\vx = \vq + \vPsi(\vq)$, where $\vPsi$ is the displacement vector.
$N$-body simulations usually end at our current time, i.e. ``today''; this is the time at which we compare them to our ML predictions.
The left panel of Fig.~\ref{fig:1} shows an example of a simulation output.

We use 210 $N$-body simulations from the Quijote suite~\cite{Villaescusa_Navarro_2020} for training (180), validation (20), and testing (10).
Each simulation contains $512^3$ particles in a box of 1 $(\mathrm{Gpc}/h)^3$ volume (nearly 5 billion light years in size). In contrast to ~\cite{He13825}, the training set data was fast approximation simulations using FastPM~\cite{fastpm} (see next section), with $32^3$ particles in a volume of 128 $(\mathrm{Mpc}/h)^3$.
The simulated universes resemble our own, and can be characterized by the value of five cosmological parameters: matter density parameter $\Omega_\mathrm{m} = 0.3175$ (this means 31.7\% of the Universe is made of dark matter), baryon density parameter $\Omega_\mathrm{b} = 0.049$, dark energy density parameter $\Omega_\Lambda = 0.6825$, Hubble parameter $h = 0.6711$, matter fluctuation within a 8 Mpc/h sphere $\sigma_8 = 0.834$, primordial spectral index $n_\mathrm{s} = 0.9624$.
In addition, we also test our model, trained on the above simulations, on three Quijote simulations with different values of the cosmological parameters, but the same configuration otherwise.

\paragraph{Fast Approximation.} $N$-body simulations can be very computationally expensive, as they typically solve the dynamics of billions of particles for thousands of time steps.
Fast approximate simulations~\cite{Tassev:2013pn, fastpm} are usually adopted when a large number of simulations are needed.
These methods save computation time by only integrating tens of time steps; they are thus less accurate than full $N$-body simulations.
We aim to build an NN model that is more accurate and faster than these approximators.
In this work we use the widely used method COLA (COmoving Lagrangian Acceleration)~\cite{Tassev:2013pn}, as implemented in the publicly available package \texttt{L-PICOLA}~\cite{Howlett:2015hfa}, as a benchmark. 
COLA solves the particle motions relative to predictions by the second order Lagrangian perturbation theory~\cite{bcgs}.
We setup \texttt{L-PICOLA} using the same configurations as the full $N$-body simulations, running it for only 10 time steps.

\paragraph{Linear Theory.} Fast and accurate solutions exist when matter distribution is close to uniform, and density fluctuations are small.
For instance, at linear order, particles move along straight lines, with the distance determined by the growth function $D$ of time: $\vPsi_\mathrm{lin}(\vq, t) = D(t) / D(t_0) \vPsi_\mathrm{lin}(\vq, t_0)$.
At early times $t_0 \to 0$, such as the starting time of the simulations, $\vPsi_\mathrm{lin}(t_0)$ agrees very well with the simulation prediction for $\vPsi(t_0)$ due to uniformity. %thus can be easily computed from its initial conditions.

The linear theory prediction is a very good approximation on large scales, where density fluctuations are small.
However, on small scales, the density contrast increases drastically and structure formation becomes non-linear, limiting the validity of linear theory predictions.
Therefore, we use $\vPsi_\mathrm{lin}(\vq,t)$ as the input to our NN model, which predicts the fully non-linear target $\vPsi(\vq,t)$ given by the $N$-body simulations at the same $t$. By design, our NN will make accurate predictions on large scales.
This is true even if we test the model on universes with cosmological parameters different from the one used for training, as we show below.

\paragraph{Neural Network Model.} Both the input (linear theory) $\vPsi_\mathrm{lin}$'s and target ($N$-body simulation) $\vPsi$'s of our NN model are functions of $\vq$'s, that form a uniform grid.
So each of them is a displacement field that can be viewed as a 3D image, with three channels being the three Cartesian components of the displacement vectors.
This allows us to apply many computer vision models to our problem.

In this work we adopt a simple U-Net / V-Net~\cite{unet, vnet} type architecture similar to that in~\cite{He13825}.
The model works on 3 levels of resolution connected in a ``V'' shape by 2 downsampling layers and 2 upsampling layers, achieved by stride-2 $2^3$ convolutions and stride-\nicefrac12 $2^3$ transposed convolutions, respectively.
Blocks of 2 $3^3$ convolutions connect the input, the resampling, and the output layers.
As in V-Net, a residual connection, in this case, $1^3$ convolutions instead of identity, are added over each of these convolution blocks.
We add batch normalization after every convolution except the first one and the last two, and leaky ReLU activation with negative slope $0.01$ after every batch normalization, as well as the first and the second to last convolutions.
As in the original ResNet~\cite{resnet}, the last activation in each residual block acts after the summation.
And as in U-Net / V-Net, at all except the bottom resolution levels, the inputs to the downsampling layers are concatenated to the outputs of the upsampling layers.
All layers have a channel size of 64, except for the input and the output (3), as well as those after concatenations (128).
Finally, an important difference from the original U-Net / V-Net is that we also add the input $\vPsi_\mathrm{lin}$ directly to the output, so effectively the network is learning the corrections to match the target $\vPsi$.

Given a displacement field, we can compute the particle positions $\vx$ and calculate their density distribution, characterized by the overdensity field $\delta(\vx) \equiv n(\vx) / \bar{n} - 1$, where $n(\vx)$ is the particle number in voxel $\vx$ and $\bar{n}$ is its mean value.
See Fig.~\ref{fig:1} for an example of $1 + \delta$ at ``today''.
Because in cosmology the density field is closely related to observables, e.g.\ galaxies, we compose a loss function that involves both $\vPsi$ and $n$, and combine them as $L = \ln(L_\delta L_\vPsi^\lambda)$, with $L_\delta$ and $L_\vPsi$ being the MSE losses on $n(\vx)$ and $\vPsi(\vq)$ respectively.
We compute $n$ from $\vPsi$ using the second order B-spline kernel (known as cloud-in-cell), so that $n$ is differentiable.
By combining the two losses with logarithm rather than summation, we can ignore their absolute magnitudes and trade between their relative values.
$\lambda$ serves as a weight on this trade-off of relative losses.
Through experiments, we find that a value of $\lambda$ from 2 to 5 yields the lowest $\ln(L_\delta L_\vPsi)$, and in this work, we use $\lambda=3$.

Limited by the GPU memory, the entire input, $\vPsi_\mathrm{lin}$ ($3\times512^3$), cannot be feed all at once to the model and needs to be cropped into smaller cubes of size $3\times128^3$.
To preserve the physical translational equivariance, we do not use any padding in the $3^3$ convolutions, resulting in a smaller output than the input in spatial size.
We compensate this by periodically padding 20 voxels per side to the input cubes.
To preserve the rotational equivariance of the simulation box, we implement data augmentation that rotates and reflects the displacement fields as in Ref.~\cite{He13825}.
We use the Adam optimizer~\cite{adam} with learning rate $0.0001$, $\beta_1 = 0.9$, $\beta_2 = 0.999$, and reduce the learning rate by half when loss does not improve for 3 epochs.

\section{Results}

\paragraph{Accuracy}

We quantify model accuracy by compare the simulation power spectra with the power spectra estimated by the NN and by the benchmark method.
The density power spectrum quantifies the correlation of density fluctuations as a function of scale:
 $P_\delta(k_i) = (N_i V)^{-1} \sum_{k_i < k \leq k_{i+1}} \delta(\vk) \delta(-\vk)$, where $V$ is the volume, $\delta(\vk)$ is the Fourier transform of $\delta(\vx)$, $k_i$'s are bin edges for the wavevector $\vk$, and $N_i$ is the number of $\vk$'s falling in $(k_i, k_{i+1}]$.
The wavenumber $k \equiv \lvert \vk \rvert$ describes the scale, with low/high $k$ representing large/small scales.
The cross-power spectrum,  $\delta_\mathrm{pred}(\vk) \delta_\mathrm{true}(-\vk)$  measures the covariance between the predicted and target density fields.
We define a transfer function, $T(k)=\sqrt{P_{\rm{pred}}(k)/P_{\rm{true}}(k)}$, where $P_{\rm{pred}}(k)$ and $P_{\rm{true}}(k)$ are the power spectra of the prediction and the simulation, respectively. If $T(k)=1$, the predictions accurate capture the amplitude of the density field.  The  cross-correlation coefficient, $r$, measures the phase correlations:
$r(k)=P_{\rm{pred}\times\rm{true}}(k)/\sqrt{P_{\rm{pred}}(k)P_{\rm{true}}(k)}$, with the numerator being the  cross-power spectrum between the prediction and the simulation.
$T$ and $r$ are good estimators to quantify the accuracy of the model, because when they are both 1 the prediction and target are identical as proved in Ref.~\cite{He13825}.
Instead of $r$, we use $1 - r^2$, which gives the amount of unexplained variance between the two fields.
Similar to $\delta$, we can also compute $P(k)$, $T(k)$, and $r(k)$ for $\vPsi$ by simply replacing the scalar product with the dot vector product.

\begin{figure}[t]
	\centering
	\includegraphics[width=0.333\textwidth,valign=c]{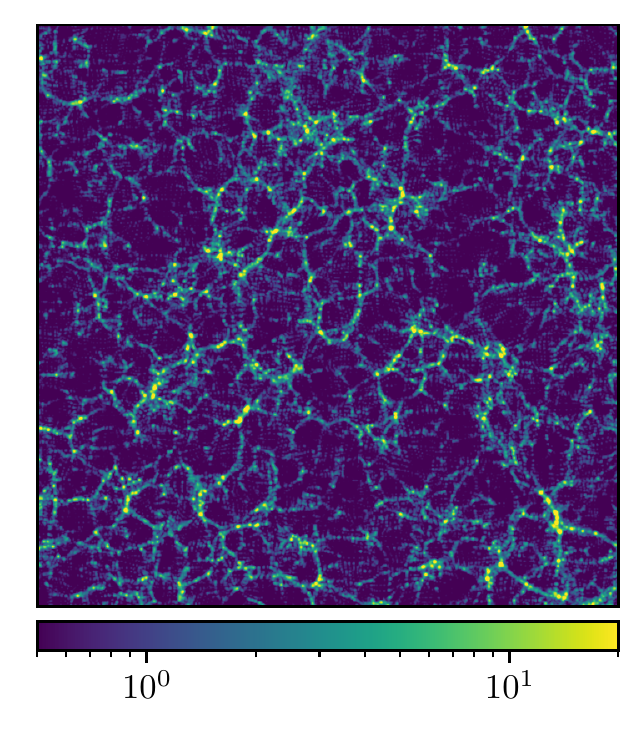}\hfill
	\includegraphics[width=0.333\textwidth,valign=c]{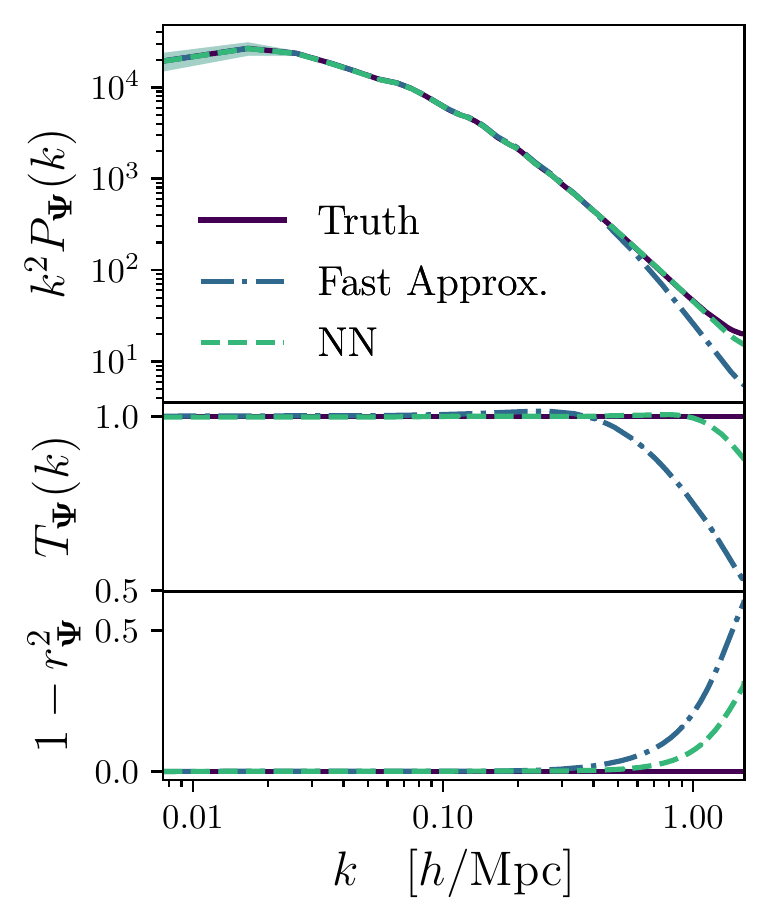}\hfill
	\includegraphics[width=0.333\textwidth,valign=c]{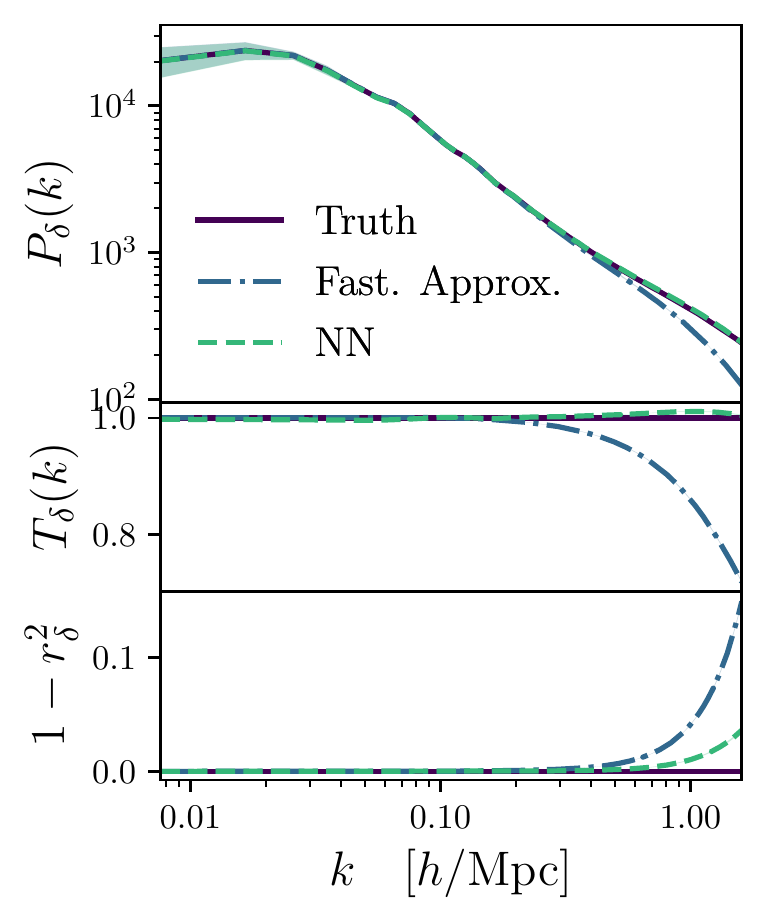}
	\caption{The left panel shows an example of the projected density field in a Universe as predicted by an $N$-body simulation.	We quantify the accuracy of the NN against the fast approximator and the N-body simulation using the power spectrum of the displacement field (middle panel) and the density constrast (right panel).
	The NN model outperforms the fast simulator in all cases.
	}
	\label{fig:1}
\end{figure}

We consider ten Quijote simulations from the test set (see the left panel of Fig.~\ref{fig:1} for an example), together with their fast approximate and NN counterparts.
From each of them, we compute their displacement and density power spectra, as well as the corresponding $T$ and $1 - r^2$.
We show the results in the central and right panels of Fig.~\ref{fig:1}.
We find that the NN outperforms the benchmark model in accuracy on all scales and for all considered quantities.
We emphasize that our NN model produces percent-level accurate results down to scales as small as $k\simeq1~h/{\rm Mpc}$.

\paragraph{Generalization to Different Universes.}

\begin{figure}[t]
    \centering
	\includegraphics[width = 0.333\textwidth]{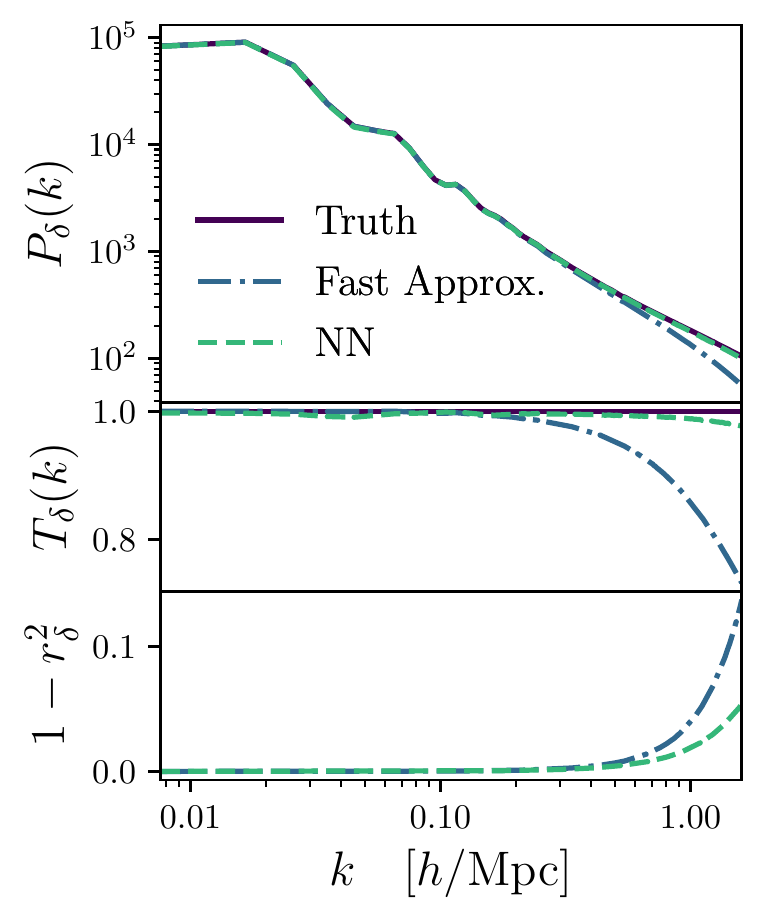}\hfill
	\includegraphics[width = 0.333\textwidth]{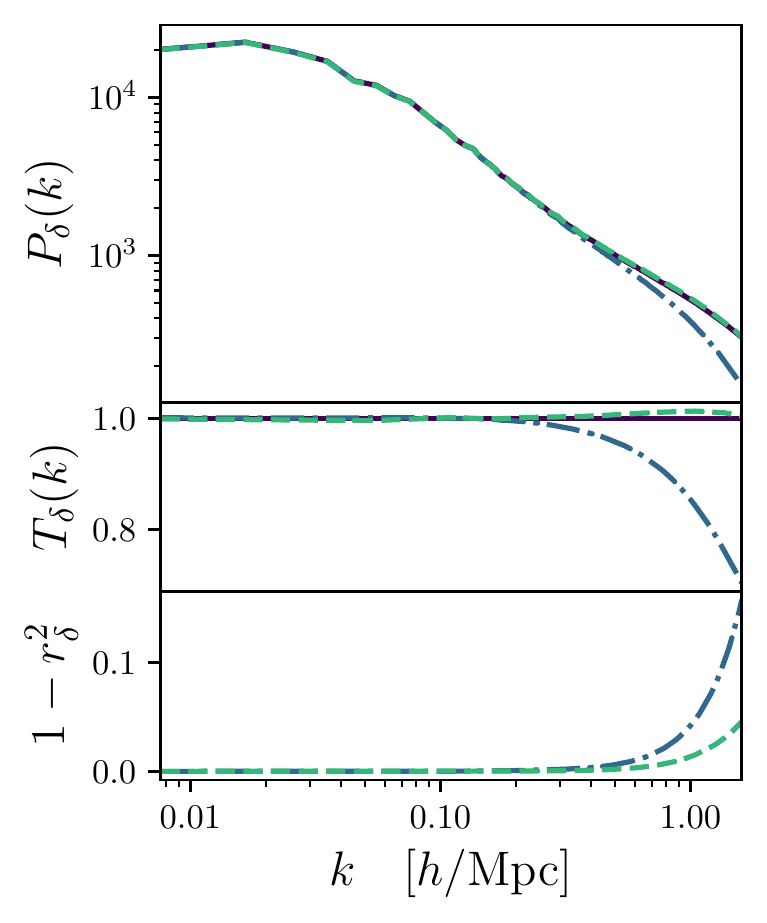}\hfill
	\includegraphics[width = 0.333\textwidth]{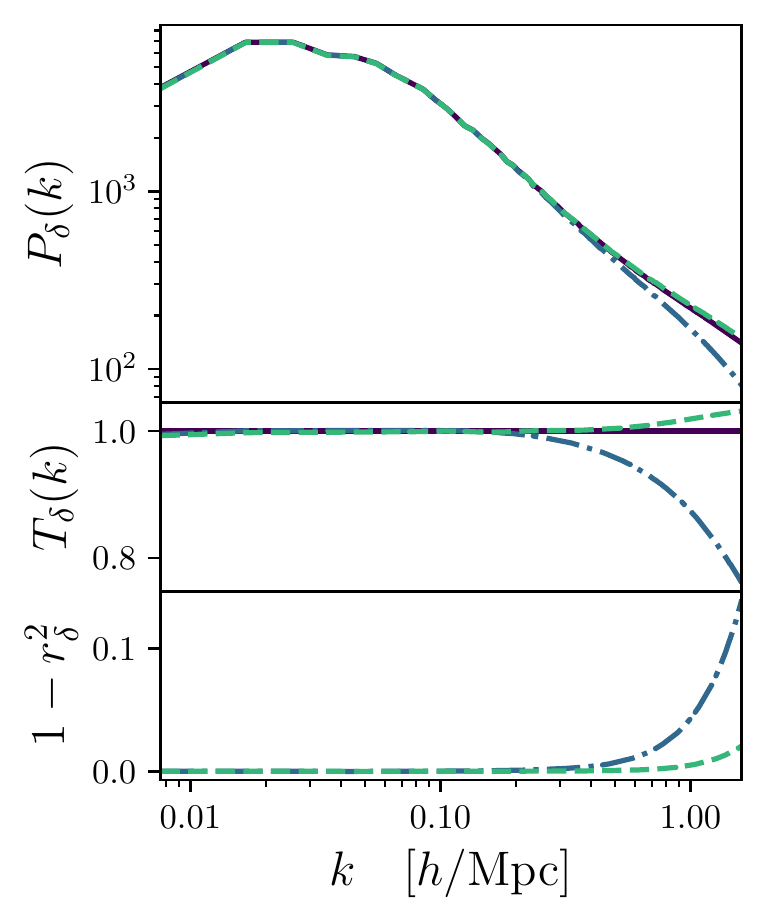}
	\caption{Accuracy comparison between predictions by the fast approximator (blue dot-dashed) and our NN (green dashed) for universes with cosmological parameters very different to those used in training: $\{\Omega_{\rm m}, \Omega_{\rm b}, h, n_s, \sigma_8\}$ equal to $\{0.1755, 0.0668, 0.7737, 0.8849, 0.6641\}$ (left), $\{0.3889, 0.0363, 0.6275, 0.9513, 0.9137\}$ (middle), $\{0.4291, 0.0337, 0.5511, 1.1687, 0.6619\}$ (right). This demonstrates that our NN generalizes very well.
}
	\label{fig:2}
\end{figure}

We test the NN model on 2000 universes with drastically different cosmological parameters, where structure formation proceeds in a very different way from that in the training simulations. In particular, we varied all five cosmological parameters: $\{\Omega_{\rm m}, \Omega_{\rm b}, h, n_s, \sigma_8\}$ that are relevant to the simulations.
We show the prediction comparisons for the density field in Fig.~\ref{fig:2}, and find that the NN generalizes very well for cosmologies with low $\Omega_\mathrm{b}/\Omega_\mathrm{m}$ and achieves a similar accuracy to that in Fig.~\ref{fig:1} (same conclusion holds for the displacement field).
It also outperforms the fast approximator COLA, even though COLA depends explicitly on the varied cosmological parameters.
This result is in concordance with that in Ref.~\cite{He13825}, where $\Omega_\mathrm{m}$ and $\sigma_8$ were varied individually. 

\paragraph{Efficiency.} We now compare the computational cost of the NN versus the fast approximation benchmark, as well as that of the $N$-body simulations.
We show in Table~\ref{tab:Bench} the total inference time to obtain the whole $512^3$ displacement field using 64 $128^3$ cubes.

The fast approximation uses 20 CPU cores (Intel \texttt{Xeon} CPU E5-2640 v4), and the NN inference uses \texttt{PyTorch}~\cite{pytorch} on 1 GPU (NVIDIA \texttt{Tesla} P100-PCIE-16GB).
The $N$-body simulation requires 500 CPU hours on 48 cores, or $\sim10^5$ seconds if we scale it to 20 cores to compare with the fast approximation.
Our NN model is two times faster than the fast approximation benchmark, and more than $10^3$ times faster than the $N$-body simulation.

\begin{table}[h]
\centering
\caption{Runtime benchmark. The NN model is $2\times$ faster than the fast approximator, and more than $10^3\times$ faster than the $N$-body simulation.}
\label{tab:Bench}
\begin{tabular}{cccc}\hline
                       & \textsc{Quijote} & $\texttt{L-PICOLA}$ & $\texttt{PyTorch-GPU}$ \\\hline
%Memory Average (GB)    & XXX      & 12                  & 28                     \\
Wall Time (s) & $10^5$      & 112                 & 59 \\\hline
\end{tabular}
\end{table}

\section{Conclusions}

In this paper, we have shown that neural networks are able to accurately emulate the output of expensive N-body simulations. We have shown that our model reproduces the results of full $N$-body simulations down to scales as small as $k=1~h/{\rm Mpc}$ at present time to within $\sim1\%$ accuracy. When compared to the state-of-the-art N-body simulation approximator in cosmology, our V-Net based network achieves better accuracy in less time. 
Furthermore, we have demonstrated that our network generalizes extremely well, by showing a comparable level of accuracy for universes that it has not been trained on.
Our method represents a big step forward in the direction of reducing the computational time needed to provide theory predictions in the non-linear regime. This paves the way to maximizing the scientific return of current and upcoming billion dollar astronomical missions.

\section*{Broader Impact}

We think the ML methods used in this work may be useful to motivate and enlighten students from different fields, especially those who are interested in astronomy, and to use new tools to help us for a better understanding of our Universe. Moreover, because of the upcoming astronomical probes for the next ten years, we believe that the outcomes of this work might help in future scientists run fast simulations of universes with high precision using a regular computer.

\begin{ack}

RAO thanks Coordenação de Aperfeiçoamento de Pessoal de Nível Superior (CAPES) and Simons Foundation.
YL thanks the Simons Foundation for support of fellowship.
The Flatiron Institute is supported by the Simons Foundation.
We acknowledge that our work was partly performed using the Princeton Research Computing resources at Princeton University which is consortium of groups led by the Princeton Institute for Computational Science and Engineering (PICSciE) and Office of Information Technology's Research Computing, and made use of the CHE cluster, managed and funded by the COSMO/CBPF/MCTI, with financial support from FINEP and FAPERJ, and operating at Javier Magnin Computing Center/CBPF. 

\end{ack}

\bibliographystyle{unsrt}
\bibliography{main}

\begin{thebibliography}{10}

\bibitem{Chang_19}
ChangHoon {Hahn}, Villaescusa-Navarro {Francisco}, Castorina {Emanuele}, and
  Scoccimarro {Roman}.
\newblock {Constraining $M_\nu$ with the Bispectrum I: Breaking Parameter
  Degeneracies}.
\newblock {\em arXiv e-prints}, page arXiv:1909.11107, Sep 2019.

\bibitem{Massara_2020}
Elena {Massara}, Francisco {Villaescusa-Navarro}, Shirley {Ho}, Neal {Dalal},
  and David~N. {Spergel}.
\newblock {Using the Marked Power Spectrum to Detect the Signature of Neutrinos
  in Large-Scale Structure}.
\newblock {\em arXiv e-prints}, page arXiv:2001.11024, January 2020.

\bibitem{Cora_19}
Cora {Uhlemann}, Oliver {Friedrich}, Francisco {Villaescusa-Navarro}, Arka
  {Banerjee}, and Sand~rine {Codis}.
\newblock {Fisher for complements: Extracting cosmology and neutrino mass from
  the counts-in-cells PDF}.
\newblock {\em arXiv e-prints}, page arXiv:1911.11158, Nov 2019.

\bibitem{Allys_2020}
E.~{Allys}, T.~{Marchand}, J.~F. {Cardoso}, F.~{Villaescusa-Navarro}, S.~{Ho},
  and S.~{Mallat}.
\newblock {New Interpretable Statistics for Large Scale Structure Analysis and
  Generation}.
\newblock {\em arXiv e-prints}, page arXiv:2006.06298, June 2020.

\bibitem{Arka_Tom_2020}
Arka {Banerjee} and Tom {Abel}.
\newblock {Nearest Neighbor distributions: new statistical measures for
  cosmological clustering}.
\newblock {\em arXiv e-prints}, page arXiv:2007.13342, July 2020.

\bibitem{He13825}
Siyu He, Yin Li, Yu~Feng, Shirley Ho, Siamak Ravanbakhsh, Wei Chen, and
  Barnab{\'a}s P{\'o}czos.
\newblock Learning to predict the cosmological structure formation.
\newblock {\em Proceedings of the National Academy of Sciences},
  116(28):13825--13832, 2019.

\bibitem{Villaescusa_Navarro_2020}
Francisco Villaescusa-Navarro, ChangHoon Hahn, Elena Massara, Arka Banerjee,
  Ana~Maria Delgado, Doogesh~Kodi Ramanah, Tom Charnock, Elena Giusarma, Yin
  Li, Erwan Allys, Antoine Brochard, Cora Uhlemann, Chi-Ting Chiang, Siyu He,
  Alice Pisani, Andrej Obuljen, Yu~Feng, Emanuele Castorina, Gabriella
  Contardo, Christina~D. Kreisch, Andrina Nicola, Justin Alsing, Roman
  Scoccimarro, Licia Verde, Matteo Viel, Shirley Ho, Stephane Mallat, Benjamin
  Wandelt, and David~N. Spergel.
\newblock The quijote simulations.
\newblock {\em The Astrophysical Journal Supplement Series}, 250(1):2, aug
  2020.

\bibitem{fastpm}
Yu~Feng, Man-Yat Chu, Uro{\v{s}} Seljak, and Patrick McDonald.
\newblock Fastpm: a new scheme for fast simulations of dark matter and haloes.
\newblock {\em Monthly Notices of the Royal Astronomical Society},
  463(3):2273--2286, 2016.

\bibitem{Tassev:2013pn}
Svetlin Tassev, Matias Zaldarriaga, and Daniel Eisenstein.
\newblock {Solving Large Scale Structure in Ten Easy Steps with COLA}.
\newblock {\em JCAP}, 06:036, 2013.

\bibitem{Howlett:2015hfa}
Cullan Howlett, Marc Manera, and Will~J. Percival.
\newblock {L-PICOLA: A parallel code for fast dark matter simulation}.
\newblock {\em Astron. Comput.}, 12:109--126, 2015.

\bibitem{bcgs}
Francis Bernardeau, S~Colombi, E~Gaztanaga, and R~Scoccimarro.
\newblock Large-scale structure of the universe and cosmological perturbation
  theory.
\newblock {\em Physics reports}, 367(1-3):1--248, 2002.

\bibitem{unet}
Olaf Ronneberger, Philipp Fischer, and Thomas Brox.
\newblock U-{Net}: {Convolutional} {Networks} for {Biomedical} {Image}
  {Segmentation}.
\newblock In Nassir Navab, Joachim Hornegger, William~M. Wells, and
  Alejandro~F. Frangi, editors, {\em Medical {Image} {Computing} and
  {Computer}-{Assisted} {Intervention} – {MICCAI} 2015}, Lecture {Notes} in
  {Computer} {Science}, pages 234--241, Cham, 2015. Springer International
  Publishing.

\bibitem{vnet}
Fausto Milletari, Nassir Navab, and Seyed-Ahmad Ahmadi.
\newblock V-{Net}: {Fully} {Convolutional} {Neural} {Networks} for {Volumetric}
  {Medical} {Image} {Segmentation}.
\newblock In {\em 2016 {Fourth} {International} {Conference} on {3D} {Vision}
  ({3DV})}, pages 565--571, October 2016.

\bibitem{resnet}
Kaiming He, Xiangyu Zhang, Shaoqing Ren, and Jian Sun.
\newblock Deep residual learning for image recognition.
\newblock In {\em Proceedings of the IEEE conference on computer vision and
  pattern recognition}, pages 770--778, 2016.

\bibitem{adam}
Diederik~P. Kingma and Jimmy Ba.
\newblock Adam: {A} {Method} for {Stochastic} {Optimization}.
\newblock In Yoshua Bengio and Yann LeCun, editors, {\em 3rd {International}
  {Conference} on {Learning} {Representations}, {ICLR} 2015, {San} {Diego},
  {CA}, {USA}, {May} 7-9, 2015, {Conference} {Track} {Proceedings}}, 2015.
\newblock arXiv: 1412.6980.

\bibitem{pytorch}
Adam Paszke, Sam Gross, Francisco Massa, Adam Lerer, James Bradbury, Gregory
  Chanan, Trevor Killeen, Zeming Lin, Natalia Gimelshein, Luca Antiga, Alban
  Desmaison, Andreas Kopf, Edward Yang, Zachary DeVito, Martin Raison, Alykhan
  Tejani, Sasank Chilamkurthy, Benoit Steiner, Lu~Fang, Junjie Bai, and Soumith
  Chintala.
\newblock {PyTorch}: {An} {Imperative} {Style}, {High}-{Performance} {Deep}
  {Learning} {Library}.
\newblock In H.~Wallach, H.~Larochelle, A.~Beygelzimer, F.~d'Alch\'{e} Buc,
  E.~Fox, and R.~Garnett, editors, {\em Advances in {Neural} {Information}
  {Processing} {Systems} 32}, pages 8026--8037. Curran Associates, Inc., 2019.

\end{thebibliography}

\end{document}